# Wide-range Angle-sensitive Plasmonic Color Printing on Lossy-Resonator Substrates

*Sarah N. Chowdhury,[1] Jeffrey Simon,[1] Michał P. Nowak,[1,2,3] Karthik Pagadala,[1] Piotr Nyga,[1,2*] Colton Fruhling,[1] Esteban Garcia Bravo,[4] Sebastian Maćkowski,[3] Vladimir M. Shalaev,[1] Alexander V. Kildishev,[1*] Alexandra Boltasseva[1*]*

Sarah N. Chowdhury, Jeffrey Simon, Michał P. Nowak, Karthik Pagadala, Piotr Nyga, Colton Fruhling, Vladimir M. Shalaev, Alexander V. Kildishev, Alexandra Boltasseva
Elmore Family School of Electrical & Computer Engineering and Birck Nanotechnology Center, Purdue University, 1205 W State St, West Lafayette, IN 47906, USA
aeb@purdue.edu and kildishev@purdue.edu

Michał P. Nowak, Piotr Nyga
Institute of Optoelectronics, Military University of Technology, 2 Kaliskiego St, Warsaw, 00-908 Poland
piotr.nyga@wat.edu.pl

Michał P. Nowak, Sebastian Maćkowski
Institute of Physics, Faculty of Physics, Astronomy, and Informatics, Nicolaus Copernicus University, Toruń 87-100, Poland

Esteban Garcia Bravo
Department of Computer Graphics Technology, Purdue University, 401 N. Grant St, Knoy Hall, West Lafayette, IN 47907, USA

We demonstrate a sustainable, lithography-free process for generating non fading plasmonic colors with a prototype device that produces a wide range of vivid colors in red, green, and blue (RGB) ([0-1], [0-1], [0-1]) color space from violet (0.7, 0.72, 1) to blue (0.31, 0.80, 1) and from green (0.84, 1, 0.58) to orange (1, 0.58, 0.46). The proposed color-printing device architecture integrates a semi-transparent random metal film (RMF) with a metal back mirror to create a lossy asymmetric Fabry-Pérot resonator. This device geometry allows for advanced control of the observed color through the five-degree multiplexing (RGB color space, angle, and polarization sensitivity). An extended color palette is then obtained through photomodification process and localized heating of the RMF layer under various femtosecond laser illumination conditions at the wavelengths of 400 nm and 800 nm. Colorful design samples with total areas up to 10 mm$^2$ and 100 μm resolution are printed on 300-nm-thick films to demonstrate macroscopic high-resolution color generation. The proposed printing approach can be extended to other applications including laser marking, anti-counterfeiting and chromo-encryption.

## 1. Introduction

Plasmonic nanostructures have been in vogue for centuries due to their remarkable ability to produce vibrant colors that depend on the nanostructure composition, morphology, geometry as well as the position of the illuminating light source. Exemplary illustrations are famous historical artifacts such as the Roman Lycurgus Cup[1] and medieval stained-glass windows.[2] The Roman Lycurgus Cup, for example, exhibits dichroic behavior – appearing ruby-red if the external illumination reflects off the surface, or jade-green if the illumination source is behind the antique. These exquisite demonstrations inspired researchers to explore the feasibility of this method for generating fade-free and environment-friendly colors instead of bleaching dyes and toxic pigments.[3] Earlier works on plasmonic color generation were performed utilizing different fabrication methods[4–6] including electron-beam lithography,[7–15] ion milling,[9,16–18] and nanoimprint lithography.[19] Although these approaches enable subwavelength resolution printing and dynamic tunability, they largely rely on expensive fabrication methodologies and are hardly industrially scalable. In this regard, contemporary studies on lithography-free optical absorbers,[20,21] resonant cavities,[22–24] thin-film multi-layered structures,[25–27] and metal-dielectric composites[28,29] have gained momentum due to their scalability, sustainability, and low cost. In particular, multilayer stacks that generate Fabry-Pérot (F-P)-like resonances have been demonstrated as narrow bandwidth and saturated color filters across a broad spectral range.[20,24,30–37] Unfortunately, the filtered or reflected colors that the multilayer stacks generate are tailored by changing the materials or layer thickness of the structure and do not allow for any adjustment or tuning post-fabrication. For applications that require a wider gamut of colors, designing and fabricating a new structure for each corresponding color becomes a time-consuming and costly process.

Recently, semi-transparent random metal films (RMFs), extensively studied as surface-enhanced Raman spectroscopy substrates,[38–40] have been employed for color printing applications.[41–43] These semicontinuous metal films absorb in a broad spectrum of light due to their random morphology and fractal-like nanostructures,[44] as these clustered nanostructures have a non-uniform absorption, contributing to significant inhomogeneous broadening.[42] RMFs can concentrate the electromagnetic energy in small, nanometer-scale voids in the films known as 'hotspots',[45] where local fields are significantly enhanced compared to the incident light.[46–49] Previously, it has been demonstrated that laser post-processing with variable laser pulse duration and wavelengths can modify the morphology of the thermally sensitive RMF layer through local heating around the nanostructures, selective melting, and fragmentation.[42–44,50–53] In this work, we achieve tunable plasmonic color printing via photomodification of an RMF-containing reflective asymmetric subwavelength F-P resonant structure, which supports interference effects that strongly depend on the angle of incidence (AOI). Through the laser photomodification process of the RMF layer, we induce spectral and polarization-sensitive changes to the reflected optical spectra, enabling the same initial device to produce a wide range of optical responses. Hence, instead of tuning the color by tailoring the structure, laser color printing tunes the color via photomodification, requiring only the fabrication of a common multilayer structure, for achieving variable colors. We also addressed how the observed colors depend on various standard illuminants (e.g., Illuminant A and D65).[54] We show that the colors are stable over a substantial time (4 months) for the as-deposited non-modified and photomodified structures. The uniformity measurements of the studied RMFs ensure an invariable macroscopic deposition for large-scale laser patterning and printing.

In summary, we obtained a five-degree multiplexing (spatial dimension RGB color space, angle, and polarization selectivity) by photomodification of the originally fabricated device to achieve a broader range of hue compared to conventional cascaded structures. The overall structure thickness is on the order of hundreds of nanometers (~300 nm) and could be manufactured on any substrate acting as a good heat sink. This approach can generate diverse color hues in a fade-free, lithography-free, and environment-friendly way. Femtosecond laser photomodification can be applied to achieve high-resolution printing using a beam size of 100 µm with a spacing of 50 µm for $\lambda = 400$ nm photomodification wavelength, comparable to the standard 300 DPI high-resolution printing. Finally, to ensure the entire approach's practicality, we present plasmonic color-printed decorative fine art produced using our in-house laser photomodification setup.

The paper is organized as follows. First, we overview and optimize the studied structure in Section 2. The optimized structures post-processed with laser photomodification for a wider gamut of colors, along with color-printed artistic images, are discussed in Section 3. The concluding remarks are addressed in Section 4 and the experimental processes are discussed in Section 5.

## 2. Asymmetric lossy resonator

The proposed structure consists of a lossy F-P resonator (**Figure 1a**) with a thin, semi-transparent lossy metal (Ag) top layer (RMF) above the percolation threshold,[42] $t_{Ag} = 20\ nm$. The Ag RMF thickness is chosen to make it robust to degradation or oxidation and to achieve a broadband optical response.[55] Under the top layer, a wavelength-scale-thick dielectric (SiO$_2$) layer is incorporated, followed by a 100-nm-thick silver mirror deposited on a glass substrate forming the lossy F-P resonator structure. Hence, the F-P like modes are formed with phase accumulations through multi-pass circulation within the dielectric layer, and the colors are observed in the reflection mode. Moreover, this architecture with Ag RMF layer adjusts the spectral width of the interference dips of the F-P cavity modes due to non-trivial phase shifts added with discontinuity and randomness.[46,56,57] A 5-nm-thin Ti layer is used as adhesion between the Ag mirror and the glass substrate.

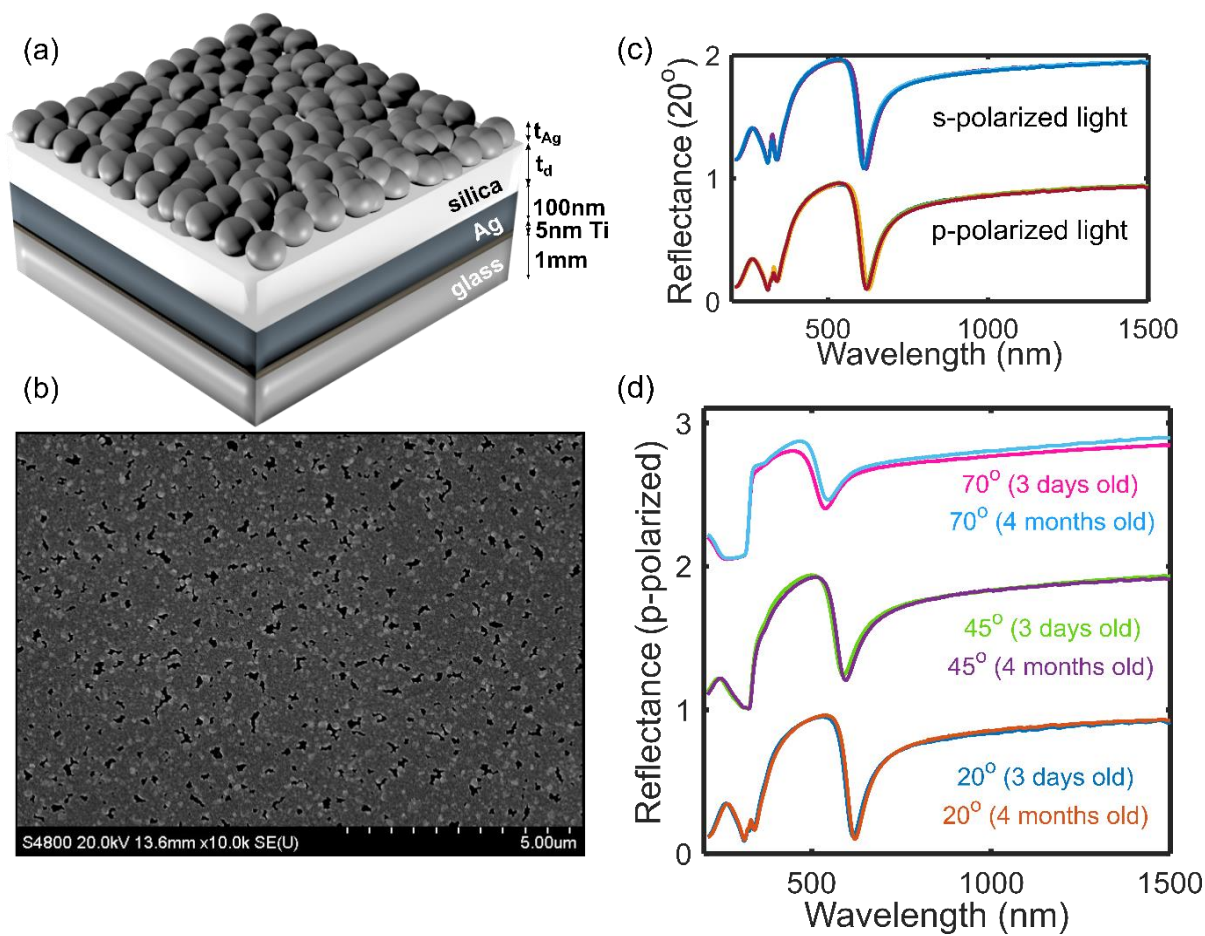

Figure 1: (a) Schematic of a lossy F-P resonator. The multilayer stack consists of a 100-nm-thick Ag layer on a glass substrate with a 5-nm-thick Ti adhesion layer. The thicknesses of the SiO$_2$ spacer and top lossy layer are $t_d$

snd $t_{Ag}$, respectively. The thickness of the substrate (1 mm) is scaled to fit the figure. (b) SEM image of the top lossy RMF layer. (c) Reflectance spectra for *p*- and *s*-polarized light at $\theta_i = 20°$ for four different sample spots. The observed near-perfect overlap proves uniformity of the RMFs for large-area printing. (d) Comparison of the optical properties of the structure 4 months apart at three different angles of incidence ($\theta_i = 20°, 45°, 70°$) for *p*-polarized beam. This confirms the long lifetime and stability of the spectral response for large-area patterning.

Scanning electron microscope (SEM) image of this RMF layer (**Figure 1b**) shows a discontinuous random metal film with voids to have a lossy metallic layer. To verify the uniformity and stability of the as-deposited RMF film, we measure the reflectance spectra of the sample at four different spots (**Figure 1c**) and at a relative time span (**Figure 1d**). A slight variation of the sample at $\theta_i = 70°$ could be attributed to the imperfections of the reference measurements based on the specific Si calibration, which is more sensitive to larger angles. The reflectance spectra of the sample at the four spots for other angles and comparison of the *s*-polarized incident light at different AOI and periods are presented in Supplementary Information (**Figures S1** and **S2**). Upon observation of the spectral uniformity and stability (Figure 1c and 1d), we conclude further that $t_{Ag} = 20\ nm$, with RMF above the percolation threshold, demonstrates better stability compared to thinner Ag films.[42,52,55] For an F-P-like multi-layer resonator, an incident light undergoes multiple passes with the interference dip's spectral location, mainly depending on the thickness of the dielectric.[58] Moreover, the broadening of the dip and the resonance quality factor depends on the contribution from random morphologies of the top RMF layer. As discussed earlier, the dielectric cavity layer ($t_d$) plays a crucial role for achieving a wide range of optical spectra sensitive to the angle of incidence. To experimentally demonstrate the interference and identify the region of interest within the visible spectrum in our study, the dielectric layer thickness was varied from 50 nm to 500 nm. **Figure 2** shows that the number of interference dips increases with the increasing dielectric thickness $t_d$, which is a characteristic effect in F-P resonators. In previous studies we determined that a high-intensity laser impinging on a discontinuous RMF surface thermally melts, modifies, and changes the morphology of the nanoparticle film areas.[42,44,49,50,52,53,59–61] Such local photomodification results in spectrally and polarization-selective changes in the scattering, transmittance, reflectance, and absorption spectra due to the gradual structural modifications occurring in the nanometer-scale areas.[38,42–44,51–53,62,63] A laser scanning setup with linearly polarized femtosecond laser pulses (repetition rate 1 kHz, pulse duration 100 fs) and operating wavelengths both at $\lambda = 800\ nm$ and $\lambda = 400\ nm$ is used for photomodifications of the studied structure. The setup generates ultrashort femtosecond laser pulses with a bandwidth of $\Delta v = 10\ nm$. The experimental study of the multilayer stack with different $t_d$ indicates that a silica thickness of $t_d = 150\ nm$ enables photomodification with femtosecond pulses both at $\lambda = 800\ nm$ and $\lambda = 400\ nm$. Indeed, Figure 2 shows that such a structure has an absorption band tail near our photomodification laser wavelengths of 400 and 800 nm. The overlap region with the photomodification wavelength allows for spectral reshaping near the absorption henceforth.

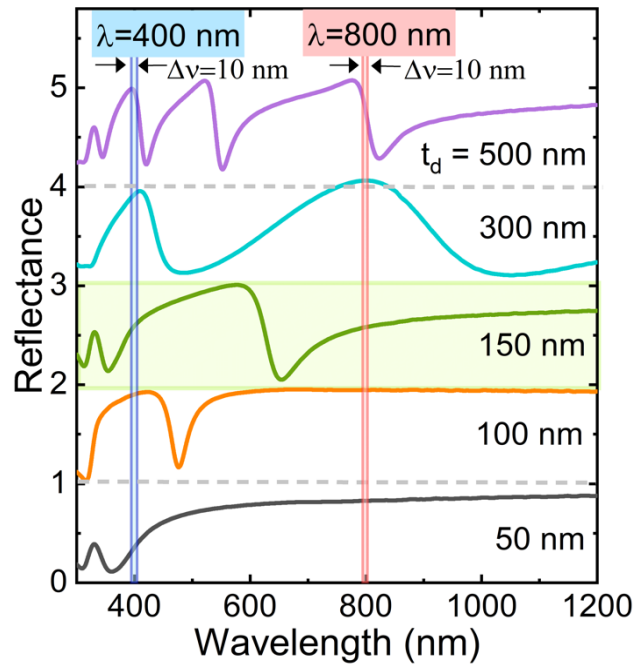

Figure 2: Reflectance spectra of different lossy resonator structures. Dielectric thickness ($t_d$) varied from 50 to 500 nm, where $t_d = 150\ nm$ shows interference dips within the region of interest. The AOI is $\theta_i = 8°$, as measured with the spectrometer. The laser photomodification wavelength with bandwidth, $\Delta\nu = 10\ nm$ is indicated by vertical lines with the selected region. Each reflectance spectrum corresponding to the dielectric thickness is shifted and normalized in the plot.

In the case of the F-P interference effects, variation and spectral reshaping in the reflection and transmission beams occur due to multiple passes and the phase change of the beam within the cavity. Hence, depending on the AOI of the beam, we observe variations in the optical distance traveled by the beam and polarization dependence (**Figure S3** Supplementary Information). This shows the tunability of colors based on the characteristics of the incident beam of light.

Another critical feature of the observed colors from this structure is the type of standard illuminants[64] that is used to record images under different illumination settings. Although there are numerous standard light sources defined by "The International Commission on Illumination (CIE)", in this paper, we consider the effect of two well-known illuminants: Illuminant A, which is the spectral distribution of incandescent light with a correlated color temperature of 2856 K and Illuminant D65, which is an average daylight (temperature 6504 K including ultraviolet wavelength region)[54] (Figure 3ab). In Figure S3 (Supplementary Information), we show that the reflectance spectra obtained for a variation of AOI, polarization state, and the kind of illuminant results in RGB values spanning the "The International Commission on Illumination (CIE)" color map. This demonstrates the tailorability of specific functionalities of the studied device depending on the illumination conditions.

**3. Laser photomodifications and plasmonic color printing**

Laser post-processing of the as-deposited sample is then conducted by varying the laser power density incident on the sample from 1.34 to 2.88 mJcm$^{-2}$ at the photomodification wavelength of $\lambda = 800\ nm$ with linearly polarized light. Details of the raster-scanning setup are presented in the Methods section. The reflectance spectra of the laser photomodified areas are shown in Supplementary Information (**Figure S4**), where distinct changes are seen with the increase of the laser intensity. This change results in different colors generated under various illuminants. The stability studies of the laser-modified areas (Supplementary Information, **Figure S5**), reveal invariance in the optical spectra within several months at room temperature and atmospheric pressure. This experiment confirms that our non-modified and photomodified structures and their colors are robust and fade-free. We visualize the range of colors of the studied optical device from the CIE color map (**Figure 3c** and **Figure 3d**) which is obtained from the measured reflectance spectra of the modified areas (Figure S4). For Illuminant A, RGB colors span from green (0.84, 1, 0.58) to orange (1, 0.58, 0.46) through pink (1, 0.92, 0.62) (**Figure 3e**). The RGB colors obtained for Illuminant D65 belong to a completely different hue from violet (0.7, 0.72, 1) to blue (0.31, 0.80, 1) (Figure 3e). Hence, we observe the dependence of the observed color on the angle of incidence of light as well as its polarization state under different illumination conditions. The damage threshold for the structure is around 4 mJcm$^{-2}$, where the photomodification has completely removed the top RMF layer and has no angle dependence of the reflected color. Colors obtained for laser photomodification at 400 nm wavelength with linearly polarized light for both Illuminant A and D65 by varying the laser power density from 0.64 to 2.64 mJcm$^{-2}$ are shown in **Figure S6** (Supplementary Information). The significance of the observed color change is illustrated and analyzed in **Figure 4**, showing the SEM images for the high and low power intensities with photomodification wavelengths of $\lambda = 800\ nm$ and $\lambda = 400\ nm$. During the photomodification process, the discontinuous surface generally fragments into smaller nanoparticles and gradually turns into spheroids as the laser fluence increases.[44] The effect is more prominent for $\lambda = 400\ nm$ where the surface quickly changes to droplet-like morphology, and a near-white color (RGB: 1, 0.99, 0.97) is observed for Illuminant D65 (Figure S6). Figure 4 also shows the almost complete removal of the entire RMF layer with $\lambda = 400\ nm$ at a higher power due to known particle delamination and laser cleaning effects.[65–67] This extreme morphological change agrees with the observed spectral reflectance resembling the bulk behavior with only a dielectric-coated Ag bottom mirror layer (Figure S6). Higher laser power density at $\lambda = 400\ nm$ has a strong effect due to its faster aggregation to smaller spheroids where higher energy is transferred from the incident beam to the RMF layer. Moreover, with this resonator, there is a shallow reflectance (~10%) around 351 nm (Figure 2), which makes the pulse at $\lambda= 400\ nm$ more efficiently absorbed by the RMF.[66] For $\lambda = 800\ nm$ the change is more gradual, starting from the longer wavelength (Figure 2), and showing gradual thermal accumulation to spheroids as the laser fluence increases (Figure 4). The morphological change of the surface induced by $\lambda = 400\ nm$ and $800\ nm$ can also be described through the optical penetration depth at the photomodification wavelength.[68] For $\lambda = 400\ nm$, the penetration depth is small (~28.3 nm),[68] which is comparable to the RMF thickness ($t_{Ag} = 20\ nm$). Thus, laser pulses are absorbed mainly by the top layer, thermally ablating both along lateral and vertical directions. Hence, with a moderate increment in the fluence, the top layer is sintered and delaminated rapidly, revealing the underlying layer (Figure 4). For $\lambda = 800\ nm$, the optical penetration depth is around 171 nm,[68] indicating the laser passes through the RMF layer. As a result, light is absorbed volumetrically, extending along a narrow vertical region towards the bottom reflector layer. This ensures that when the laser beam passes through the resonant structure, it does not have any pronounced heating, diffusing laterally along the surface, and that heating is confined within a relatively small region.

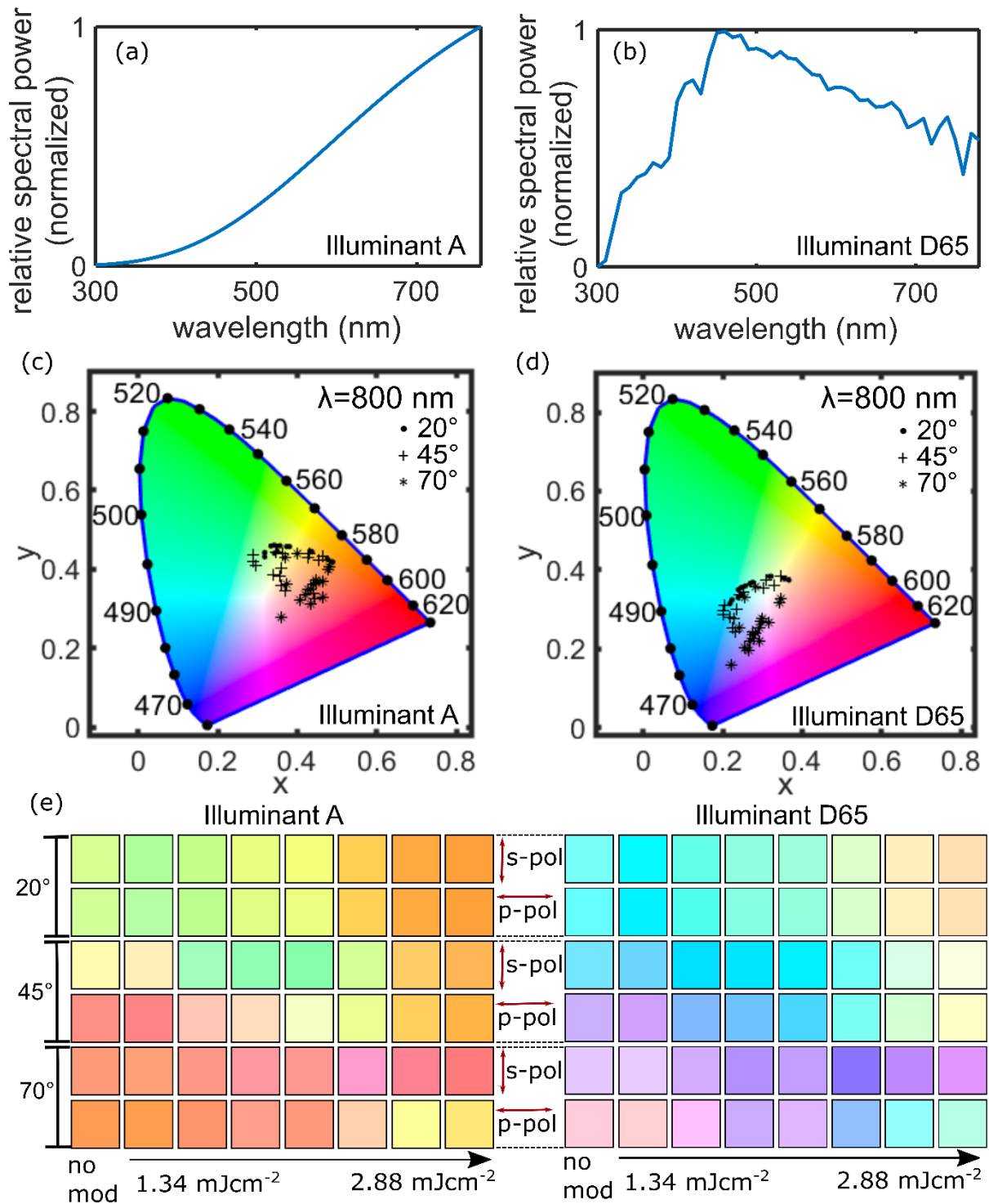

Figure 3: Spectral power distribution of (a) CIE Illuminant A and (b) D65. The International Commission on Illumination (CIE) 1931 chromaticity diagrams with (c) Illuminant A and (d) D65 are shown to have different sets of colors with varying laser power and AOI of $\theta_i = 20°, 45°, 70°$. The colors generated for non-modified (no mod) and laser photomodified samples under Illuminant A and D65 show angular and polarization dependence. Laser power density is varied from 1.34 to 2.88 mJcm$^{-2}$ for operating photomodification wavelength $\lambda = 800\ nm$ with linearly polarized light, and s and p polarization are denoted by *s*-pol and *p*-pol for each set of AOI. RGB colors with (e) Illuminant A and D65 show a broad range for angular dispersion, polarization, and laser photomodifications.

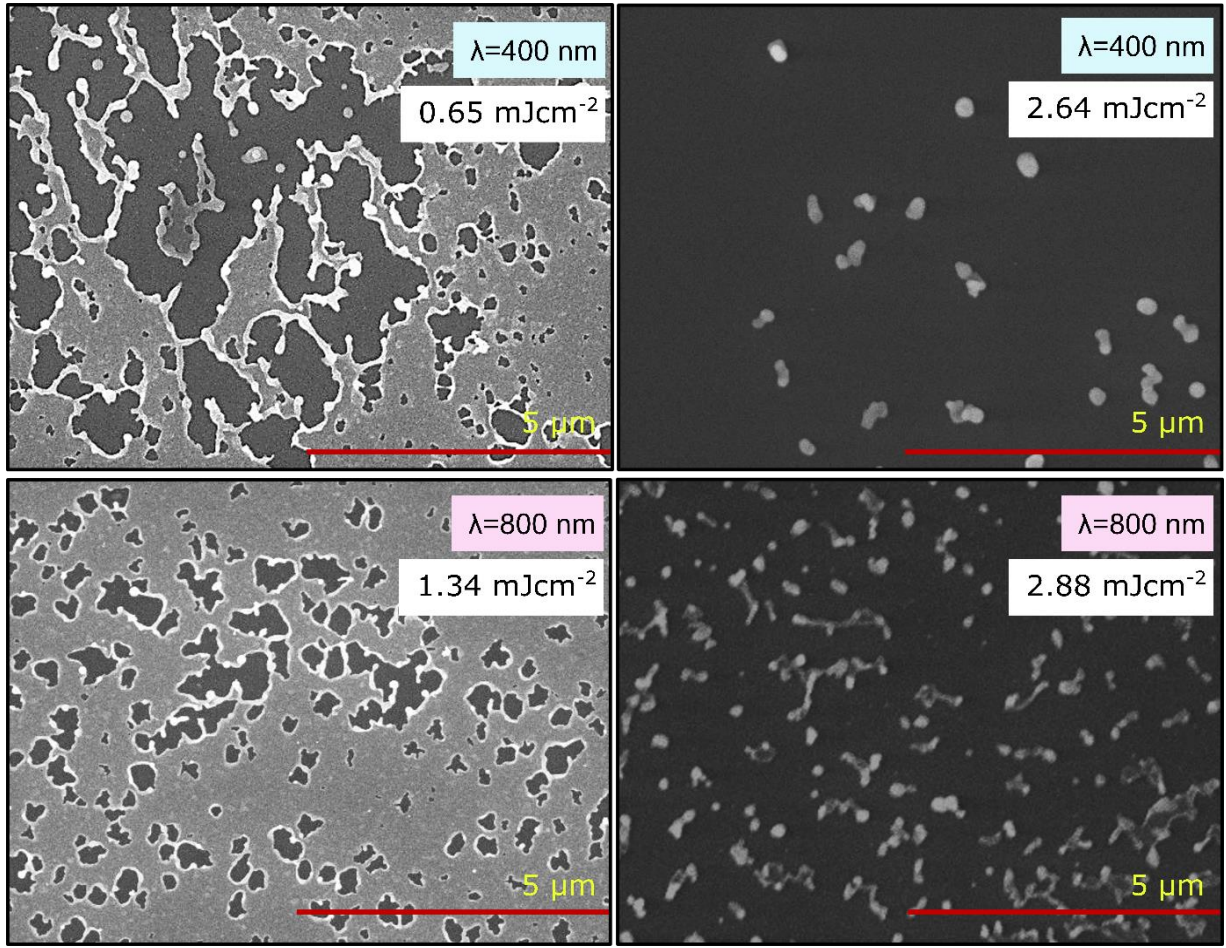

Figure 4: SEM images of laser-modified areas with laser photomodification wavelengths of $\lambda = 400\ nm$ (0.65 mJcm$^{-2}$ and 2.64 mJcm$^{-2}$) (top panel) and $\lambda = 800\ nm$ (1.34 mJcm$^{-2}$ and 2.88 mJcm$^{-2}$), respectively (bottom panel). We observe the evolution and morphological change with the increased laser power intensity. The surface is thermally ablated at higher laser power, with gradual delamination and dewetting to a droplet-like structure.

Figure 3e shows the colors of non-modified and photomodified surfaces with illumination from both *s*-polarized and *p*-polarized light. With larger AOI, there is a significant difference in the mapped colors obtained from reflected *s*- and *p*-polarized light. Correspondingly, the change in the polarization state of light with rendered colors is more evident for the photomodification wavelength $\lambda = 800\ nm$. For the non-modified sample at $\theta_i = 45°$ and a wavelength of 550 nm, the structure reflects the *p*-polarized light and absorbs (resonates) the *s*-polarized light (**Figure 5**). For the sample at the same incident angle and a wavelength of 588 nm, the structure absorbs (resonates) the *p*-polarized light and reflects the *s*-polarized light. This effect demonstrates that as the light polarization is switched from *s*-polarized to *p*-polarized, there is a redshift in the resonance of the structure. Thus, within the 477 – 674 nm spectral range, polarization-switchable reflectivity occurs at two corresponding wavelengths. After photomodification with a higher laser intensity (2.88 mJcm$^{-2}$) of $\lambda = 800\ nm$ light, this effect becomes very broad throughout the entire visible spectrum. Note that from 418 to 677 nm the sample is spectrally reflecting a dominant *s*-polarized light compared to the *p*-polarized light. This relationship interchanges in the near infra-red regime. A similar phenomenon happens for the non-modified sample at $\theta_i = 70°$ within the range of 436 to 563 nm (i.e., within the green colors), where the resonance behavior altogether blue shifts when the light polarization is switched from *s*-polarized to *p*-polarized. This polarization dependence originates from the strong correlation of the near-field anisotropy effects of the connected nanoparticles in the RMF layer along the incident light polarization direction.[69] When light travels through the island-like RMF layer, it encounters multiple scattering and extinction events along the beam path. In contrast with normal incidence, such multiple scattering is more prominent at larger incident angles, where significant scattering effects arise from the discontinuity along the traverse direction. Hence, this discontinuity arising from complex Ag aggregate topologies introduces near-field scattering along the beam path, changing its polarization at the detector. At the detector, where the polarization state of light (*s* or *p*-polarized), is recorded, a blue shift of the resonance peak (either *s* or *p*) for $\theta_i = 70°$ compared to $\theta_i = 45°$ is observed. The depolarization factor causes different interactions for *s*- and *p*- polarized light depending on the component of the electric field along the plane. The island-like cluster in the RMF layer has a statistical superposition of Ag nanoparticles, each with a different depolarization factor. In general, the shape and orientation of any inclusion or nanoparticle can be assigned to a depolarization value or geometric factor where the value for uniform spherical inclusions is $L = \frac{1}{3}$.[70] This has a direct relationship with the nanoparticle polarizability. For shapes that deviate from spherical symmetry, the polarizabilities correlate with the propagation direction of the incident light. Due to this correlation, the anisotropic depolarization factor, similar to that of prolate or oblate spheroids, becomes dependent on the ratio of the particle principal axes (depolarization factor, $L \neq \frac{1}{3}$) and results in the shift of the observed peaks.[70] Therefore, in contrast with the case of normal incidence, the polarized white light beam at $\theta_i = 70°$ encounters more Ag particles resembling oblate spheroid shapes, leading to increased depolarization factors and shifting the averaged peak to a shorter wavelength. For the photomodified areas, a similar angular effect takes place. However, in this case, the resonance and reflectance dip gradually disappear due to local photomodification, partially removing the RMF layer. This modification then results in a reflection predominantly from the bottom Ag layer. Hence, although in the case of photomodification, the effect of scattering from large clusters or spheroids can be attributed to the change in polarization, the underlying layers play the key role in the observed spectra. For photomodified areas, the direction of laser modification or raster scanning is parallel to the photomodification beam polarization. Such a striking change in polarization gives us a wide range of dramatically different colors and can also switch polarization through angular dispersion and morphological evolution[71] due to laser photomodification (Figure 4). The effect could be applied to anti-counterfeiting applications and laser marking beyond the visible range due to changes induced in the near-infrared spectral range. Specifically, the next-generation color/visual security labels can be equipped with an additional polarization-detection authentication enabling a more advanced and sophisticated security system.

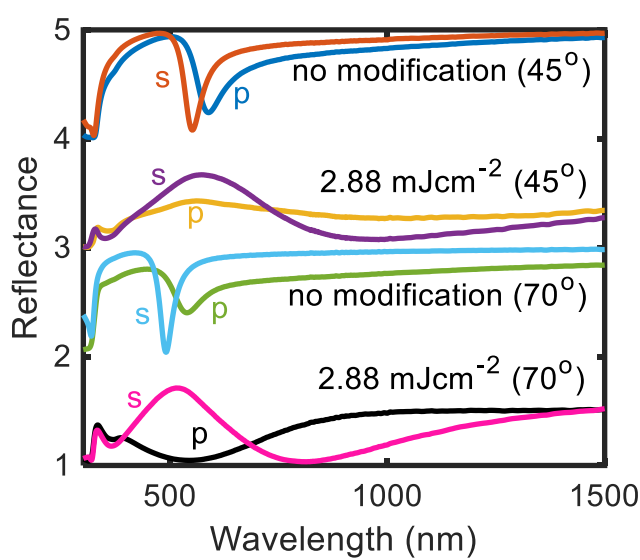

Figure 5: Reflectance spectra of non-modified sample and laser photomodified (2.88 mJcm$^{-2}$) with photomodification wavelength, λ = 800 nm for ($\theta_i$ = 45°) and ($\theta_i$ = 70°). A change and shift in resonance are observed from observing *s*-polarized to *p*-polarized light and vice versa.

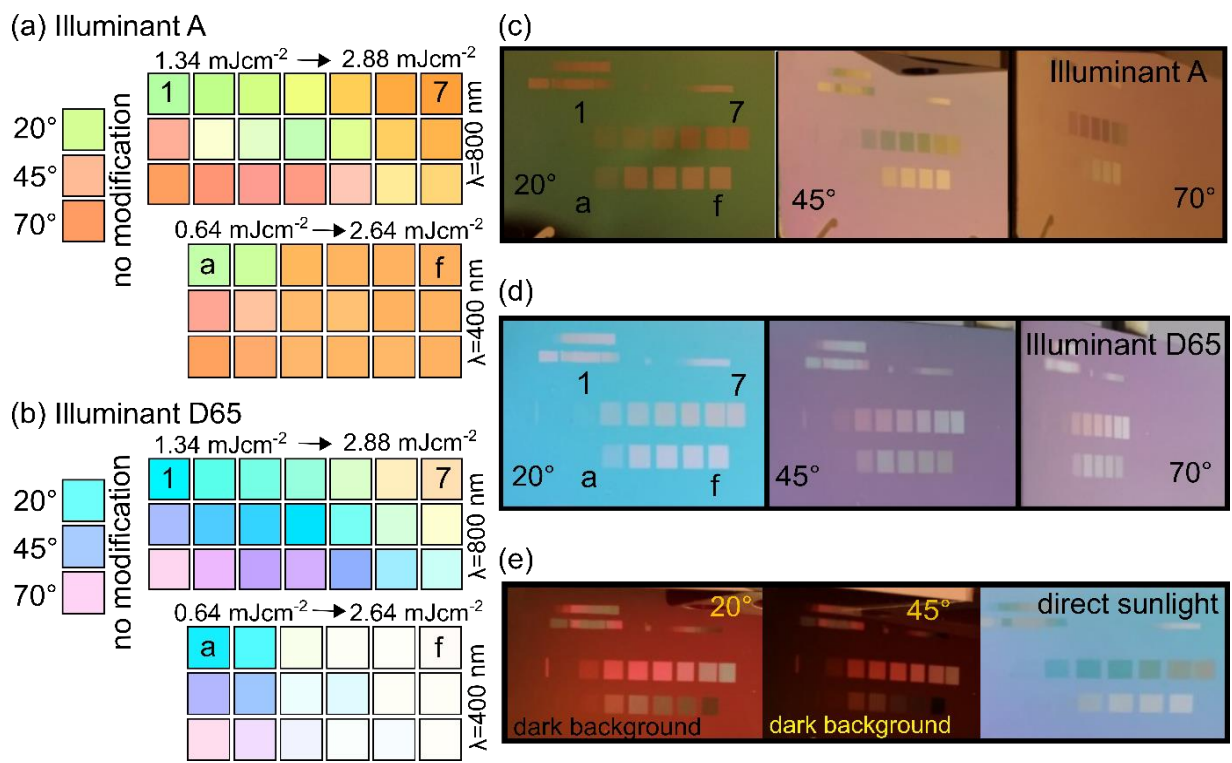

Figure 6: The colors generated under Illuminants (a) A and (b) D65 showing angular dependence for the non-modified and laser photomodified areas. Laser power density varies from 1.34 to 2.88 mJcm$^{-2}$ for the photomodification wavelength $\lambda$ = 800 nm and from 0.64 to 2.64 mJcm$^{-2}$ for $\lambda$ = 400 nm with linearly polarized light. RGB colors are obtained by averaging the two polarizations. Optical images of the laser-modified areas for Illuminants (c) A and (d) D65, when the sample is oriented at different angles respectively. The corresponding laser photomodified areas are marked 1-7 (all the subsequent columns) for $\lambda$ = 800 nm and a-f for $\lambda$ = 400 nm. (e) Images of the photomodified areas taken with dark background and under direct sunlight.

To show the applicability of the developed structure, we record optical images of the photomodified areas with unpolarized light under Illuminants A and D65. **Figures 6a** and **6b** show the RGB colors obtained from Figure 3, after averaging the two polarizations (*s* and *p*) to obtain the effect of unpolarized light. The optical images are recorded with a Fujifilm X-T200 camera (sRGB (standard RGB), ISO-8000, no flash) **(Figures 6c** and **6d)**. The recorded images and the photomodified colors under different illumination conditions demonstrate the generation of a broad spectrum of visible colors from blue to green through yellow and violet to blue. We also recorded optical images of the non-modified and photomodified areas with a dark background and under direct sunlight (**Figure 6e**). The former is taken in a relatively dark background coming from the camera surface (Fujifilm X-T200) mounted in front of the sample with ISO-8000 and color representation mapping of sRGB. The direct sunlight color reflects the natural light where the sample is placed horizontally outdoors and is not calibrated.

Fine-art samples, achieved with this coloring technique, are fabricated by controlling the actuators of an XYZ motorized stage capable of raster scanning (**Figure 7** with more details in Methods section). For each image, an initial calibration swatch, created by varying the laser power while color printing sequential squares, determined the power settings for each printed color. Multilayer bitmap images specified the location of each color on the sample (Supplementary Information, **Figure S7**). A Python script controlled the instrumentation for each power setting to print a single-layer pattern. Repeated printing of different image layers at different powers resulted in a multicolored final printed image. **Figure 7a** and **Figure 7b** show the printed images where the sample is oriented at different angles and under variable illumination intensity as well as sources. For Figure 7a, high and low illumination intensity is used where, high, and low intensity means the light source is placed in front of the sample for high intensity and at an inclined angle for low intensity. For **Figure 7c**, we see images under different light settings. Variation of intensity for artificial light is achieved by moving the light source gradually away from the sample. Natural daylight intensity variation is achieved by putting the sample directly or away from the sunlight. This setting is not calibrated. The recorded images hence demonstrate high-resolution color printing comparable to standard coloring techniques.

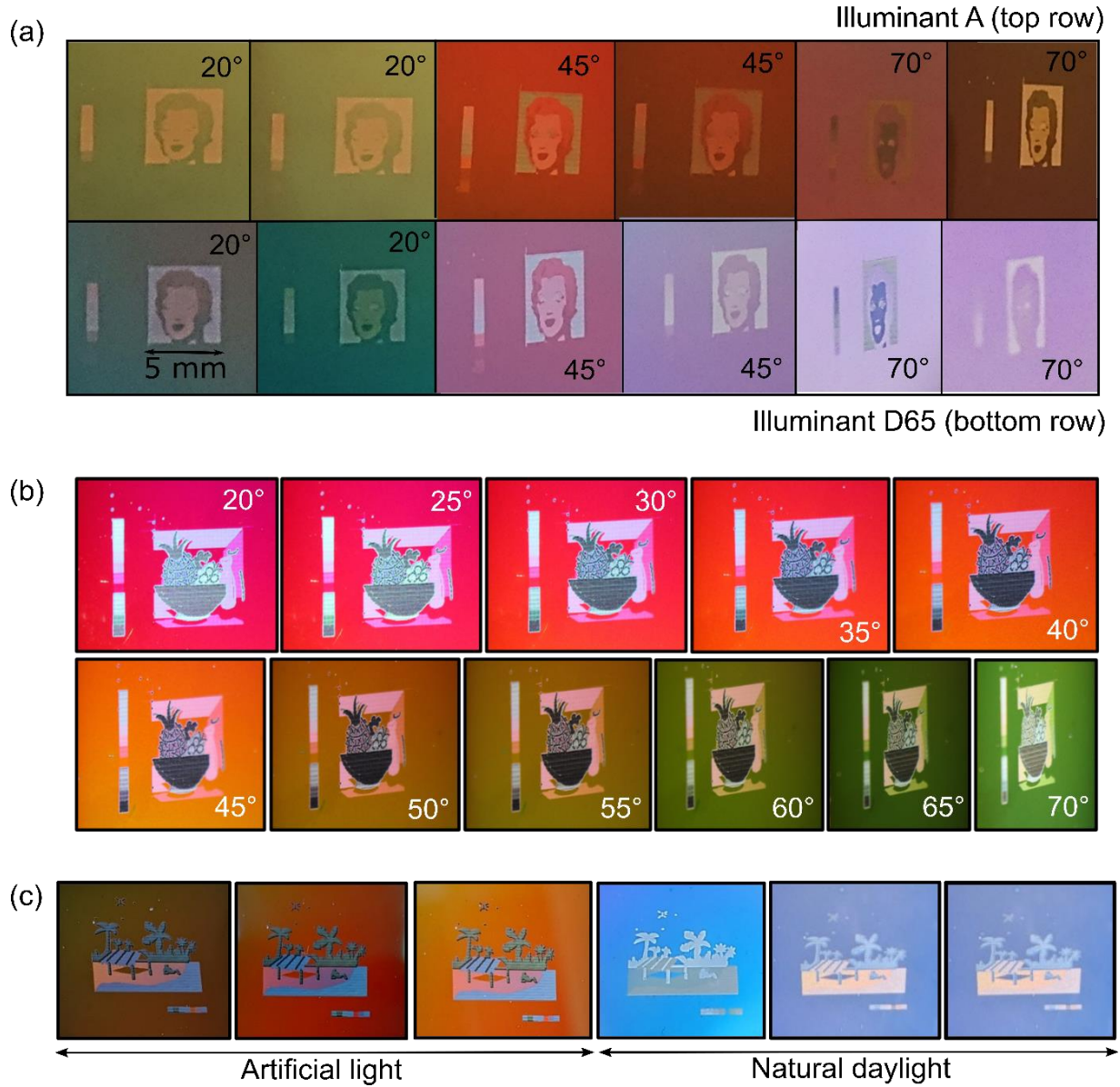

Figure 7: (a) Optical images of Andy Warhol's Marilyn Monroe shown under Illuminants A (top row) and D65 (bottom). Top row: The first two are at sample orientation relative to the camera at 20° for high and low Illuminant A intensity, the next two at 45°, and the last at 70°. Bottom row: The same settings for Illuminant D65. (b) "A fruit basket" image, where the sample is rotated at different angles to view colored images. The illumination used is the laboratory mixed-light setting. (c) Image of "a lazy afternoon" under artificial light sources like a lamp and natural daylight illumination with different intensities. The image sizes in (b) and (c) are 6.5 mm$^2$, and 10.75 mm$^2$ respectively. Images in (a) and (c) are taken with Fujifilm X-T200 (sRGB, ISO-8000) and (b) with Canon EOS 77D (sRGB, ISO-1250).

## 4. Conclusion

We report on scalable, low-cost, fade-free plasmonic color printing in multilayer samples with a lossy resonator architecture that utilizes semi-transparent random metal films (RMFs) and Fabry-Pérot (F-P)-like geometry. We demonstrate multiplexing of optical spectra and the reflected color through RGB color space, adjustable angular and polarization sensitivities. This results in a palette of colors that was further expanded with femtosecond laser post-processing with a beam resolution of 100 μm. Stability of the as-deposited and laser post-processed colors is studied for obtaining uniform, robust, and non-fading colors. A wide gamut of colors from green to yellow and from violet to blue is demonstrated, where a desired color range can be tailored to a given illuminant. Such capabilities with a single versatile optical device can be further applied for optical data storage, polarization switching, and anti-counterfeiting applications. Moreover, the macroscopic uniformity tested on the developed samples can be utilized for practical applications of disordered metasurfaces and meta-optics. Development of the in-house laser setup by optimizing laser parameters such as line spacing, scanning speed, exposure time, and beam size can enable advanced artistic designs to reach the resolution of lithography-based color printing. Finally, this optimal configuration with the raster scanning laser setup can be utilized as a standard coloring technique, paving the way for full-color printing on miniaturized and macroscopic substrates.

## 5. Methods

*Fabrication of a lossy resonator:* The lossy resonator formed from a lossy Ag layer, silica spacer, and a silver mirror deposited on a glass substrate was fabricated in a single process using an electron-beam physical vapor deposition (PVD) technique. The glass substrates were pre-cleaned with Piranha solution (3 parts $H_2SO_4$:1 part $H_2O_2$) for 15 minutes and thoroughly rinsed with distilled water. After drying out with nitrogen gas, the substrates were sonicated in solvents (toluene, acetone, and isopropyl alcohol) and dried thoroughly.[48] Next, a titanium adhesion layer, silver mirror, silica spacer, and lossy Ag layer were deposited in a high-vacuum deposition chamber (PVD Products, Inc.), base pressure $3.33\times10^{-6}$ mbar at room temperature. Silicon dioxide ($SiO_2$, 99.99% purity), titanium (Ti, 99.99% purity), and silver (Ag, 99.99% purity) from the Kurt J. Lesker Company were used for fabricating all structures. The deposition rate (1 Å/s for all materials) and layer thickness were monitored with a quartz crystal microbalance.

*Laser post-processing and design setup:* Laser photomodification of the lossy resonator was performed in ambient conditions using 800 nm femtosecond pulses generated by a Ti:Sapphire femtosecond seed laser and ultrafast amplifier (Spectra-Physics, Solstice Ace;1 kHz, 100 fs, 800 nm, linear polarization). To perform photomodification at 400 nm, an inserted second harmonic generation (SHG) crystal doubled the frequency of the original femtosecond pulse. A TTL shutter (Vincent Associates Uniblitz Optical Shutter) controlled the number of pulses for each photomodification event. A Variable ND Filter controlled the pulse power, and thus, the color resulting from the selective modification of the sample. The laser beam was focused using a single lens and the 1e$^{-2}$ Gaussian beam size was determined using the knife-edge technique. The beam size calculated for $\lambda = 800\ nm$ is 300 µm and $\lambda = 400\ nm$ is 100 µm. To print areas of uniform color, samples were mounted on a motorized XYZ stage (Zaber Technologies Inc.) capable of raster scanning and controlled with a computer interface. To ensure uniformity of modification over the large area, we use a 50 µm X and Y-axis (raster) step. A Python-generated code for instrumentation control patterned various designs onto the samples. A digital photography camera captured the color images of the printed structures at multiple angles while a rotation stage precisely controlled the position of the sample. **Figure S7** (Supplementary Information) presents a schematic of the setup.

*Sample characterization:* A field emission scanning electron microscope (FESEM, Hitachi S-4800) was used to characterize the nanostructure of semi-transparent RMFs. Angle-resolved reflectance spectra were obtained using a dual-rotating compensator ellipsometer (RC2, J. A. Woollam Co., Inc.). For the laser-modified spots, focusing probes were used to limit the beam to the laser-modified area. Window effects were accounted for using a reference Si wafer for the reflectance baseline. Total reflectance spectra of as-fabricated and photomodified structures were measured using a spectrophotometer (Perkin Elmer, Lambda 950) equipped with an integrating sphere (150 mm) module (8° angle of incidence used for reflectance measurement) and linear polarizers for Figure 2. Spectralon was used as a reference sample for reflectance measurements. CIE 1931 color coordinates from reflectance spectra were calculated with a proprietary MATLAB script. Coordinates were computed using color matching functions, spectral power distribution of the illuminating light, and measured sample reflection in the visible spectrum. Ohta and Robertson[54] and Song et al.[5] provide more details on calculations with CIE color space equations.

**Supporting Information**

Supporting Information is available from the Wiley Online Library or the authors.

**Acknowledgements**

The Purdue team acknowledges financial support by the U.S. Office of Naval Research (ONR) under Awards N00014- 21-1-2026, and N00014-20-S-B001, and the U.S. Air Force Office of Scientific Research (AFOSR) under Award FA9550-20-1-0124 and FA9550-21-1-0299. Military University of Technology team acknowledges financial support by the Military University of Technology under grant UGB725-2022. Jeffrey Simon acknowledges support from the National Science Foundation Graduate Research Fellowship under Award DGE-1842166. This material is based upon work supported by the National Science Foundation Graduate Research Fellowship under Grant No. DGE-1842166. Any opinions, findings, conclusions, or recommendations expressed in this material are those of the authors(s) and do not necessarily reflect the views of the National Science Foundation. The authors also thank Peigang Chen for his assistance with camera images.

Received: ((will be filled in by the editorial staff))
Revised: ((will be filled in by the editorial staff))
Published online: ((will be filled in by the editorial staff))

References

[1]  I. Freestone, N. Meeks, M. Sax, C. Higgitt, *Gold Bull* **2007**, *40*, 270.
[2]  S. Pérez-Villar, J. Rubio, J. L. Oteo, *J Non Cryst Solids* **2008**, *354*, 1833.
[3]  H. Ali, *Water Air Soil Pollut* **2010**, *213*, 251.
[4]  A. Kristensen, J. K. W. Yang, S. I. Bozhevolnyi, S. Link, P. Nordlander, N. J. Halas, N. A. Mortensen, *Nat Rev Mater* **2016**, *2*, 1.
[5]  M. Song, D. Wang, S. Peana, S. Choudhury, P. Nyga, Z. A. Kudyshev, H. Yu, A. Boltasseva, V. M. Shalaev, A. V. Kildishev, *Appl Phys Rev* **2019**, *6*, 041308.
[6]  D. Wang, Z. Liu, H. Wang, M. Li, L. J. Guo, C. Zhang, *Nanophotonics* **2023**, *12*, 1019.
[7]  X. M. Goh, R. J. H. Ng, S. Wang, S. J. Tan, J. K. W. Yang, *ACS Photonics* **2016**, *3*, 1000.
[8]  K. Kumar, H. Duan, R. S. Hegde, S. C. W. Koh, J. N. Wei, J. K. W. Yang, *Nat Nanotechnol* **2012**, *7*, 557.
[9]  G. Si, Y. Zhao, J. Lv, M. Lu, F. Wang, H. Liu, N. Xiang, T. J. Huang, A. J. Danner, J. Teng, Y. J. Liu, *Nanoscale* **2013**, *5*, 6243.
[10] A. S. Roberts, A. Pors, O. Albrektsen, S. I. Bozhevolnyi, *Nano Lett* **2014**, *14*, 783.
[11] S. J. Tan, L. Zhang, D. Zhu, X. M. Goh, Y. M. Wang, K. Kumar, C.-W. Qiu, J. K. W. Yang, *Nano Lett* **2014**, *14*, 4023.
[12] R. J. H. Ng, R. V. Krishnan, H. Wang, J. K. W. Yang, *Nanophotonics* **2020**, *9*, 533.
[13] C. U. Hail, G. Schnoering, M. Damak, D. Poulikakos, H. Eghlidi, *ACS Nano* **2020**, *14*, 1783.
[14] X. Zhu, C. Vannahme, E. Højlund-Nielsen, N. A. Mortensen, A. Kristensen, *Nat Nanotechnol* **2015**, *11*, 325.
[15] H.-C. Wang, O. J. F. Martin, H.-C. Wang, O. J. F. Martin, *Adv Opt Mater* **2023**, *11*, 2202165.
[16] H. Lochbihler, *Opt Express* **2009**, *17*, 12189.
[17] T. Xu, Y.-K. Wu, X. Luo, L. J. Guo, *Nat Commun* **2010**, *1*, 1.
[18] F. Cheng, J. Gao, T. S. Luk, X. Yang, *Sci Rep* **2015**, *5*, 11045.

[19] K. Kumar, H. Duan, R. S. Hegde, S. C. W. Koh, J. N. Wei, J. K. W. Yang, *Nat Nanotechnol* **2012**, *7*, 557.
[20] F. Ding, L. Mo, J. Zhu, S. He, *Appl Phys Lett* **2015**, *106*, 061108.
[21] C. S. Park, S. S. Lee, *ACS Appl Nano Mater* **2021**, *4*, 4216.
[22] K. Mao, W. Shen, C. Yang, X. Fang, W. Yuan, Y. Zhang, X. Liu, *Sci Rep* **2016**, *6*, 1.
[23] Y. G. Kim, Y. J. Quan, M. S. Kim, Y. Cho, S. H. Ahn, *International Journal of Precision Engineering and Manufacturing - Green Technology* **2021**, *8*, 997.
[24] P. Cencillo-Abad, D. Franklin, P. Mastranzo-Ortega, J. Sanchez-Mondragon, D. Chanda, *Sci Adv* **2023**, *9*, eadf7207.
[25] K. T. Lee, D. Kang, H. J. Park, D. H. Park, S. Han, *Materials* **2019**, *12*, 1050.
[26] J. A. Dobrowolski, *Appl Opt* **1981**, *20*, 74.
[27] S. Mader, O. J. F. Martin, S. Mader, O. J. F. Martin, *Light: Advanced Manufacturing* **2021**, *2*, 385.
[28] N. Destouches, N. Sharma, M. Vangheluwe, N. Dalloz, F. Vocanson, M. Bugnet, M. Hébert, J. Siegel, *Adv Funct Mater* **2021**, *31*, 2010430.
[29] N. Dalloz, V. D. Le, M. Hebert, B. Eles, M. A. Flores Figueroa, C. Hubert, H. Ma, N. Sharma, F. Vocanson, S. Ayala, N. Destouches, *Advanced Materials* **2022**, *34*, 2104054.
[30] Z. Yang, Y. Zhou, Y. Chen, Y. Wang, P. Dai, Z. Zhang, H. Duan, *Adv Opt Mater* **2016**, *4*, 1196.
[31] M. Seo, J. Kim, H. Oh, M. Kim, I. U. Baek, K. D. Choi, J. Y. Byun, M. Lee, *Adv Opt Mater* **2019**, *7*, 1900196.
[32] J. Kim, H. Oh, M. Seo, M. Lee, *ACS Photonics* **2019**, *6*, 2342.
[33] Z. Lin, Y. Long, X. Zhu, P. Dai, F. Liu, M. Zheng, Y. Zhou, H. Duan, *Optik (Stuttg)* **2019**, *178*, 992.
[34] J. Zhao, M. Qiu, X. Yu, X. Yang, W. Jin, D. Lei, Y. Yu, *Adv Opt Mater* **2019**, *7*, 1900646.
[35] C. S. Park, S. S. Lee, *ACS Appl Nano Mater* **2021**, *4*, 4216.
[36] T. J. Palinski, T. J. Palinski, A. Tadimety, I. Trase, B. E. Vyhnalek, G. W. Hunter, E. Garmire, J. X. J. Zhang, *Opt Express* **2021**, *29*, 25000.
[37] M. A. Kats, R. Blanchard, P. Genevet, F. Capasso, *Nat Mater* **2012**, *12*, 20.
[38] M. Liszewska, B. Budner, M. Norek, B. J. Jankiewicz, P. Nyga, *Beilstein Journal of Nanotechnology* **2019**, *10*, 1048.
[39] B. N. Khlebtsov, V. A. Khanadeev, E. V. Panfilova, D. N. Bratashov, N. G. Khlebtsov, *ACS Appl Mater Interfaces* **2015**, *7*, 6518.
[40] V. P. Drachev, M. D. Thoreson, E. N. Khaliullin, V. J. Davisson, V. M. Shalaev, *Journal of Physical Chemistry B* **2004**, *108*, 18046.
[41] S. M. Novikov, C. Frydendahl, J. Beermann, V. A. Zenin, N. Stenger, V. Coello, N. A. Mortensen, S. I. Bozhevolnyi, *ACS Photonics* **2017**, *4*, 1207.
[42] P. Nyga, S. N. Chowdhury, Z. Kudyshev, M. D. Thoreson, A. V. Kildishev, V. M. Shalaev, A. Boltasseva, *Opt Mater Express* **2019**, *9*, 1528.
[43] A. S. Roberts, S. M. Novikov, Y. Yang, Y. Chen, S. Boroviks, J. Beermann, N. A. Mortensen, S. I. Bozhevolnyi, *ACS Nano* **2019**, *13*, 71.
[44] P. Nyga, V. P. Drachev, M. D. Thoreson, V. M. Shalaev, *Appl Phys B* **2008**, *93*, 59.
[45] D. P. Tsai, J. Kovacs, Z. Wang, M. Moskovits, V. M. Shalaev, J. S. Suh, R. Botet, *Phys Rev Lett* **1994**, *72*, 4149.
[46] V. M. Shalaev, *Nonlinear Optics of Random Media: Fractal Composites and Metal-Dielectric Films*, Springer, **2000**.
[47] M. D. Thoreson, J. Fang, A. V. Kildishev, L. J. Prokopeva, P. Nyga, U. K. Chettiar, V. M. Shalaev, V. P. Drachev, *J Nanophotonics* **2011**, *5*, 051513.

[48] U. K. Chettiar, P. Nyga, M. D. Thoreson, A. V Kildishev, V. P. Drachev, V. M. Shalaev, *Appl Phys B* **2010**, *100*, 159.
[49] V. P. Drachev, S. V. Perminov, S. G. Rautian, V. P. Safonov, in *Optical Properties of Nanostructured Random Media*, Springer, Berlin, Heidelberg, **2002**, pp. 115–148.
[50] S. N. Chowdhury, P. Nyga, Z. Kudyshev, E. Garcia, A. V. Kildishev, V. M. Shalaev, A. Boltasseva, in *Conference on Lasers and Electro-Optics*, Optical Society Of America, San Francisco, **2020**, p. SF2R.2.
[51] M. D. Ooms, Y. Jeyaram, D. Sinton, *Langmuir* **2015**, *31*, 5252.
[52] S. N. Chowdhury, P. Nyga, Z. A. Kudyshev, E. Garcia Bravo, A. S. Lagutchev, A. V. Kildishev, V. M. Shalaev, A. Boltasseva, *ACS Photonics* **2021**, *8*, 521.
[53] M. P. Nowak, B. Stępak, M. Pielach, Y. Stepanenko, T. Wojciechowski, B. Bartosewicz, U. Chodorow, P. Wachulak, P. Nyga, in *Nanophotonics IX*, SPIE-Intl Soc Optical Eng, **2022**, p. 90.
[54] N. Ohta, A. R. Robertson, *Colorimetry: Fundamentals and Applications* **2006**, 63.
[55] X. Wang, C. Santschi, O. J. F. Martin, *Small* **2017**, *13*, 1700044.
[56] Y. Su, X. Tang, G. Huang, P. Zhang, *Opt Commun* **2020**, *464*, 125483.
[57] D. A. Genov, A. K. Sarychev, V. M. Shalaev, *Journal of Nonlinear Optical Physics and Materials* **2003**, *12*, 419.
[58] J. M. Vaughan, *The Fabry-Perot Interferometer: History, Theory, Practice and Applications* **2017**, 1.
[59] P. Nyga, A. V. Kildishev, S. N. Chowdhury, A. Boltasseva, Z. Kudyshev, V. M. Shalaev, *Optical Device, Method of Using the Same, and Method of Making the Same*, **2020**, 16/794964.
[60] V. M. Shalaev, *Phys Rep* **1996**, *272*, 61.
[61] V. M. Shalaev, A. K. Sarychev, *Phys Rev B* **1998**, *57*, 13265.
[62] F. Meng, A. Pucci, *physica status solidi (b)* **2007**, *244*, 3739.
[63] V. M. Shalaev, R. Botet, A. V Butenko, *Phys Rev B Condens Matter* **1993**, *48*, 6662.
[64] J. Morovic, J. Lammens, *"Color Management"*, in *Colorimetry: Understanding the CIE System*, J. Wiley, **2007**.
[65] D. Qi, D. Paeng, J. Yeo, E. Kim, L. Wang, S. Chen, C. P. Grigoropoulos, *Appl Phys Lett* **2016**, *108*, 211602.
[66] Y. Oh, M. Lee, *Appl Surf Sci* **2017**, *399*, 555.
[67] F. Font, S. Afkhami, L. Kondic, *Int J Heat Mass Transf* **2017**, *113*, 237.
[68] D. Paeng, J. Yeo, D. Lee, S. J. Moon, C. P. Grigoropoulos, *Appl Phys A Mater Sci Process* **2015**, *120*, 1229.
[69] J. Laverdant, S. Buil, B. Bérini, X. Quélin, *Phys Rev B* **2008**, *77*, 165406.
[70] U. Kreibig, M. Vollmer, *Experimental Results and Discussion. In: Optical Properties of Metal Clusters*, Springer, Berlin, Heidelberg, **1995**.
[71] K. J. Berean, V. Sivan, I. Khodasevych, A. Boes, E. Della Gaspera, M. R. Field, K. Kalantar-Zadeh, A. Mitchell, G. Rosengarten, *Adv Opt Mater* **2016**, *4*, 1247.